\documentclass[letter,12pt]{article}
\usepackage{graphicx,amssymb,amsmath}

\def\beq{\begin{equation}}
\def\eeq{\end{equation}}
\def\bea{\begin{eqnarray}}
\def\eea{\end{eqnarray}}
\def\bmat{\begin{pmatrix}}
\def\emat{\end{pmatrix}}
\def\bei{\begin{itemize}}
\def\eei{\end{itemize}}
\def\Im{\, {\rm Im}}
\def\GeV{\, {\rm GeV}}
\def\TeV{\, {\rm TeV}}
\def\arg{\, {\rm arg}}
\def\ecm{\, {\rm e\, cm}}
\def\fb{\, {\rm fb}}
\newcommand{\Fig}[1]{Fig.~\ref{#1}}

\newcommand{\Eq}[1]{Eq.(\ref{#1})}

\makeatletter
\renewcommand{\section}{\@startsection{section}{1}{0em}%
        {-3.25ex \@plus -1ex \@minus -.2ex}%
        {2.0ex \@plus.2ex}%
        {\normalfont\large\bfseries}}
\renewcommand{\subsection}{\@startsection{subsection}{2}{0em}%
        {-2.75ex\@plus -1ex \@minus -.2ex}%
        {1.25ex \@plus .2ex}%
        {\normalfont\bfseries}}
\renewcommand{\subsubsection}%
        {\@startsection{subsubsection}{3}{0em}%
        {-2.0ex\@plus -1ex \@minus -.2ex}%
        {1.0ex \@plus .2ex}%
        {\normalfont\itshape}}
\makeatother

\begin{document}
\baselineskip=18pt

\begin{titlepage}

\noindent
\begin{flushright}
MCTP-08-65 \\
CERN-PH-TH-2008-229\\
\end{flushright}
\vspace{1cm}

\begin{center}
  \begin{Large}
    \begin{bf}
Comparison of electric dipole moments and \\
the Large Hadron Collider for probing \\
CP violation in triple boson vertices

        \end{bf}
  \end{Large}
\end{center}

\vspace{0.5cm}
\begin{center}
\begin{Large}
Sunghoon Jung$^a$, James D. Wells$^{b,a}$ \\
\end{Large}

\vspace{0.3cm}
\begin{it}
${}^{a}$Michigan Center for Theoretical Physics (MCTP) \\
        ~~University of Michigan, Ann Arbor, MI 48109-1120, USA \\
\vspace{0.1cm}
${}^{b}$CERN, Theory Division, CH-1211 Geneva 23, Switzerland
\vspace{0.1cm}
\end{it}

\vspace{1cm}
\end{center}

\begin{abstract}

CP violation from physics beyond the Standard Model  may reside
in triple boson vertices of the electroweak theory.
We review the effective theory description  and discuss how CP violating contributions to these vertices might be discerned by electric dipole moments (EDM) or diboson production at the Large Hadron Collider  (LHC). Despite triple boson CP violating interactions entering EDMs only at the two-loop level, we find that EDM experiments are generally more powerful than the  diboson processes. To give example to these general considerations we perform the comparison between EDMs and collider observables within  supersymmetric theories that have heavy sfermions, such that substantive EDMs at the one-loop level
are disallowed.  EDMs
generally remain more powerful probes, and next-generation EDM experiments may surpass even the most optimistic assumptions for LHC sensitivities.

\end{abstract}

\vspace{3cm}

\begin{flushleft}
\begin{small}
November 2008
\end{small}
\end{flushleft}

\end{titlepage}


\tableofcontents

\section{Introduction}

CKM phases explain all observed CP violations. However, baryogenesis apparently requires more CP-violation than is provided for by the Standard Model (SM).  Thus physics beyond SM should contain new source of CP violation that is somehow  small enough not to be in conflict with experiment.

CP violation from new physics can manifest itself in several ways. One way is by measuring an electric dipole moment (EDM) of a fermion. No EDM has been found to date.
The current experimental electron EDM (eEDM) bound is  $d_e \leq 2.14 \times 10^{-27} \ecm$ at $95\%$ CL \cite{NewEDM}, which already puts a strong constraint on physics beyond the SM. In supersymmetric theories~\cite{Martin:1997ns}, the  eEDM induced at one-loop is usually larger than this bound so we need several assumptions \cite{Abel:2001vy, Demir:2003js, Brhlik:1998zn} or cancellation mechanisms \cite{Ibrahim:1997gj, Ibrahim:1998je} to avoid this limit for a wide range of parameter space.

CP violation can also be seen in CP asymmetries of particle energy-momentum distributions at colliders. One such CP asymmetric collider observable was proposed recently using the interference effect between CP conserving and violating $WWZ$ interactions in the diboson production processes at LHC~\cite{Kumar:2008ng}. This observable may be able to improve collider sensitivities on CP violating couplings such as triple boson vertices (TBV) by up to two orders of magnitude from the most recent LEP results. Since we expect that abundant diboson production will occur at LHC, and they have clean tri-lepton decay signals, this observable is useful to probe new physics at the LHC.
This improvement raises the hope of discovery, and it is worthwhile studying the possible reach of both the collider observable and EDM measurements in more detail.

Intuition holds in the physics community that EDMs are the most powerful probes of new physics contributions to flavor-preserving CP violation.  That intuition is largely based on the varieties of  supersymmetric theories that have dipole moments induced at the one-loop level. However,  given the possibility of the LHC increasing the probing sensitivity by a few orders of magnitude,  we investigate how solid that intuition is within the context of theories that have suppressed one-loop contributions to EDMs.  Our primary example is supersymmetry with heavy sfermion masses. Ultimately, we shall not disagree that EDMs are unlikely to be supplanted by the LHC in the search for new sources of CP violation. We detail the path to that strengthened conclusion below.

\section{Triple Boson Vertices and CP Violating Observables}

\subsection{Triple Boson Vertices Effective Interactions} \label{sec:tbv}

\begin{figure}
\centering
\includegraphics[angle=0,width=0.95\textwidth]{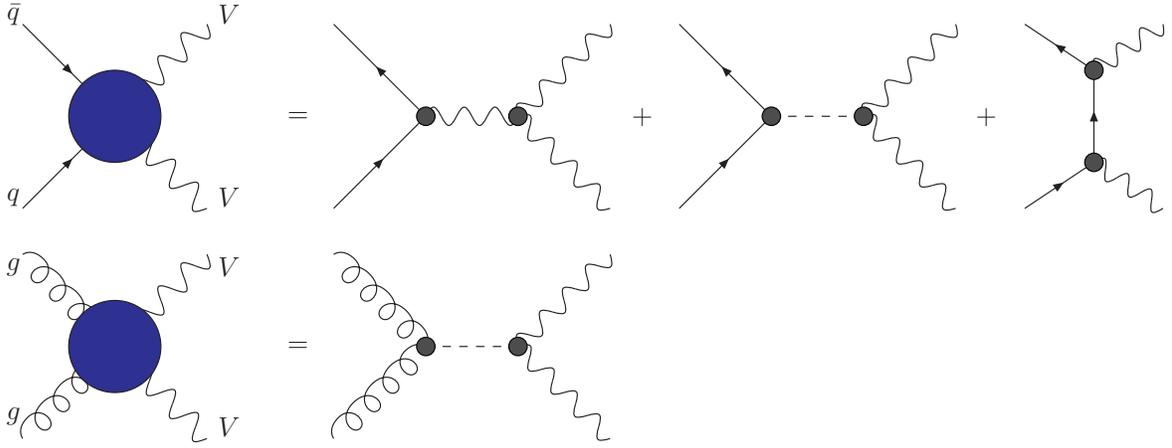}
\caption{Diboson production processes at the LHC. Blobs on the right-hand side are effective interactions in the low-energy effective theory. These effective interactions contain both CP-even and odd contributions.}\label{diboson:LHC}
\end{figure}

Diboson production channels at the LHC are described in Fig.~\ref{diboson:LHC} using the low-energy effective theory below the electroweak scale. This effective theory is obtained by integrating out heavy particles in physics beyond the SM. The modified SM interactions which now contain both CP-even and odd interactions are represented as small blobs in the figure. One can see from the figure that we should study the diboson production channels at the LHC, involving triple boson vertices (TBV) $VVV, \, hVV$ and couplings with fermions $Vff,\, hff$.

We will focus only on TBV among them. One reason for this is that we can easily extend our work to include fermion couplings without changing the conclusions. Secondly, CP-odd effective couplings are  mediated by particles in the BSM and are loop suppressed. Any charged particle couples to the vector bosons, whereas only a small number of particles couple to a specific fermion typically. Therefore, TBV is   more generally present than more direct CP-violating couplings with fermions. In addition, large CP-odd $Vff$ couplings with $V=\gamma$ can induce an EDM without loop-suppression, as is discussed in section~\ref{sec:edm}. Since other $Vff$ couplings are presumably related with the $\gamma ff$ coupling in an underlying theory, it is difficult to avoid the  experimental EDM limit with large CP-odd $Vff$ couplings. Thus, a meaningful analysis can be carried out with TBV only.

The effective Lagrangian of CP-odd TBV is~\cite{Hagiwara:1986vm}
\bea
{\cal L}_{\textrm{CP-odd TBV}} &=& ig_{WWV} \left( \widetilde{\kappa}_V W^+_\mu W_\nu \widetilde{V}^{\mu\nu} \, + \, \frac{\widetilde{\lambda}_V}{m^2_W} W^{+ \mu}_\nu W^\nu_\rho \widetilde{V}^\rho_\mu \, + \, g_4^V W^+_\mu W_\nu (\partial^\mu V^\nu + \partial^\nu V^\mu ) \right) \nonumber\\
&& \quad +\, \frac{g\, M_W}{4} \,\left( g_{H_iWW} \, \widetilde\eta_i^W \, \widetilde{W}^{\mu \nu} W^+_{\mu \nu} \,+\, g_{H_iZZ} \,\frac{\widetilde\eta_i^Z}{2c_W^2} \widetilde{Z}^{\mu \nu} Z_{\mu \nu} \, \right) H_i   \label{CPodd:WWV}
\eea
where  $g_{WW\gamma}=-e,\, g_{WWZ}=-e\cot\theta_W$, and $g_{H_iVV}$ is the ratio of CP-even $H_iVV$ coupling to SM $H_iVV$ coupling. $V$ can be $\gamma$ or $Z$. $V^{\mu\nu} = \partial^\mu V^\nu - \partial^\nu V^\mu$ and likewise for $W^{\mu\nu}$. Index $i$ runs for two light (CP-even) Higgses. $g_4^V$ is C-odd while others are P-odd, so $g_4^V$ is not relevant for our work as discussed in section~\ref{sec:coll}. Higher dimensional operators are suppressed by the electroweak scale $M_W$. Higgs couplings to photons and gluons can also be written in the same way. These effective couplings are actually momentum dependent. However, we can reasonably choose to study constant on-shell couplings as argued in Appendix A.

It is useful to know the $SU(2)\times U(1)$ invariant dimension-six operators that generate the effective triple gauge couplings in \Eq{CPodd:WWV} after electroweak symmetry breaking. $H^\dagger H V^{\mu\nu} \tilde{V}_{\mu\nu}$ and $D_\mu H^\dagger T^a D_\nu H \tilde{V}^{a \mu\nu}$ generate $\tilde\kappa_V$. $\tilde\lambda_V$ is generated by $\epsilon_{abc}\widetilde{W}^{a\mu}_\nu W^{b\nu}_\rho W^{c\rho}_\mu$ which does not involve Higgs fields. $V_{\mu\nu}$ and $W_{\mu\nu}$ here are full field strengths. CP-odd neutral $VVV$ couplings are not generated by these operators. As couplings with photons and couplings with $Z$ bosons are presumably related in an underlying theory, we shall reduce redundancy and give results in terms of the $Z$ boson coupling only.

\subsection{CP Asymmetric Collider Observable} \label{sec:coll}

CP asymmetries at colliders are observables well-known to probe CP violating
interactions~\cite{CPasymmetries}. It has been shown that if absorptive SM backgrounds are known well
the LHC may be sensitive to $\tilde{\lambda}_Z$ coupling perhaps as low as the $\tilde{\lambda}_Z \lesssim 0.001$ with $100 \fb^{-1}$, which would be a significant improvement over LEP2 capabilities, for example~\cite{Kumar:2008ng}. This sensitivity was achieved based on the fact that the cross section proportional to the $\epsilon_{\mu\nu\rho\sigma}$ tensor is a signal of the CP violation since the tensor is odd under time reversal. Thus, only $P$ and $CP$ odd couplings in Eq.~(\ref{CPodd:WWV}) are potentially able to be probed with this precision. Although no equivalent small value has been estimated for $\tilde \kappa_Z$, we shall suggest by analogy to $\tilde\lambda_Z$ that it may be possible.  The $C$-odd coupling $g_4^V$ can be probed in other ways and will not be treated in this paper.

CP violating Higgs couplings can also be probed at the LHC in the same way, in principle. Several other collider observables sensitive to Higgs couplings have been studied as well based on the angular distributions of final leptons. The sensitivities on the CP violating $hZZ$ coupling are usually expected to be around $\tilde{\eta}^Z \lesssim O(0.1) $ with $100 - 300 \fb^{-1}$ of data from the process $h \to ZZ \to 4l$ at LHC, and possibly $O(0.01)$ from Higgsstrahlung at a future $e^+ e^-$ linear collider~\cite{Godbole:2007cn}. As we study EDM sensitivities to the  CP violating couplings involving the Higgs boson, we compare results to the $\tilde{\eta}^Z \lesssim O(0.1)$ LHC expected sensitivity.

\subsection{Electric Dipole Moments} \label{sec:edm}

\begin{figure}[t]
\centering
\includegraphics[angle=0,width=0.95\textwidth]{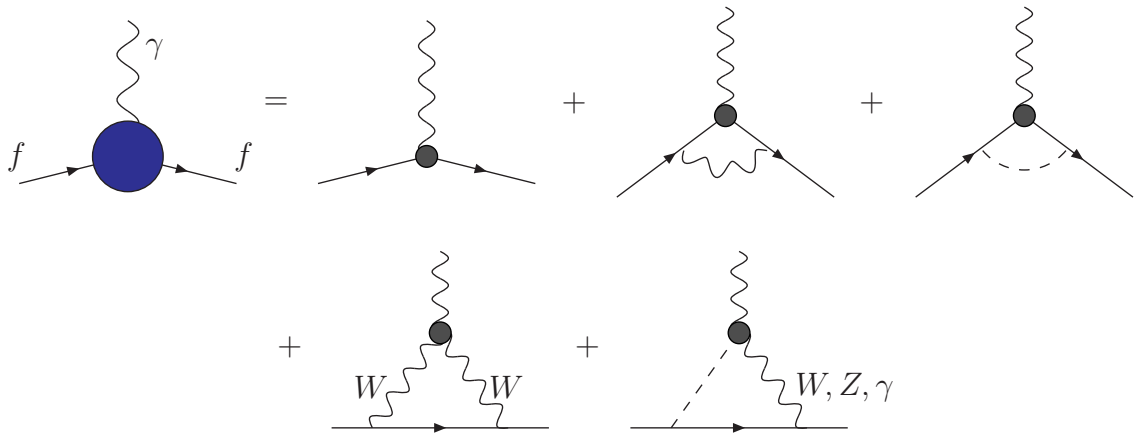}
\caption{EDM diagrams at one and two-loop orders with the effective CP-odd couplings represented as small blobs. Since effective couplings are radiatively generated, the first diagram on the right-hand side is a one-loop contribution and others are two-loop.
}\label{EDM:one-two}
\end{figure}

One and two-loop order generation of EDMs are shown in Fig.~\ref{EDM:one-two} using the effective theory. CP-odd effective couplings in the effective theory are represented as small blobs in the figure. Since effective CP-odd couplings are generated at loop order, the tree-level diagram on the right-hand side implies a one-loop contribution to the EDM and others are two-loop contributions.

We want to avoid one-loop induced EDM in our study. The first reason for this is that a one-loop induced EDM is usually larger than experimental bounds in many models. Secondly, we want to give a ``one loop advantage" to collider observables -- we expect that TBVs are generated at one-loop order, which only then enables EDMs at two-loop order. As can be seen from the figure, one-loop EDM (first diagram) corresponds to CP-odd $Vff$ couplings with $V=\gamma$ and on-shell external particles. The suppression of the one-loop induced EDM roughly implies the smallness of CP-odd $Vff$ couplings, and vice versa. Then the two-loop diagrams in the first line of Fig.~\ref{EDM:one-two} are also suppressed.

We will consider only two-loop contributions in the second line of Fig.~\ref{EDM:one-two} with specified insertions of effective couplings. We note that these effective interactions are CP violating TBVs that were necessary for the diboson production process as discussed in sec~\ref{sec:tbv}. At two-loop order, $WW$ and scalar-vector can transmit CP violation to SM fermions, whereas scalar-scalar mediation is very small due to small Yukawa couplings. Only $WW$ can mediate CP-violation without Higgs bosons because there is no CP-odd triple neutral electroweak boson couplings. EDMs are generated only through these CP violating TBVs as long as we ignore quartic and higher effective couplings. It can also be inferred that EDMs and CP violating TBVs depend on the same CP phases as will be discussed in sec~\ref{sec:CPphase}.

The electric dipole operator should be RG evolved from high scale down to the fermion mass scale at which the fermion EDM is defined. Renormalization group flow mixes this operator with other operators with the same quantum numbers such as chromo-electric dipole, three-gluon Weinberg operator~\cite{Weinberg:1989dx} and $SU(2)$ analogies of these. For electron EDM (eEDM), not all are  relevant since the electron is colorless. The remaining $SU(2)$ operators are relatively suppressed by multiple powers of $g/g_S$ and the QED renormalization effects are smaller than QCD. We will not consider renormalization effects for eEDM.

In this paper we focus on the electron EDM since the experimental measurements are excellent and improving, and the theory computation has minimal theoretical uncertainty. Of course, one expects a high degree of correlation of one EDM to other EDMs in most theories of physics beyond the SM, and later we shall briefly study the correlation of electron EDM and neutron EDM. As stated earlier, the current sensitivity limit on the eEDM is  $d_e \leq 2.14 \times 10^{-27} \ecm$ at $95\%$ CL~\cite{NewEDM}. Upon surveying the literature, one expects that the future eEDM sensitivity of the near-term future experiments to be approximately $10^{-29} \ecm$~\cite{Kawall:2003ga,Hudson:2002,Liu:edm}. When appropriate, we shall use these numbers as benchmark sensitivities in the numerical discussion ahead.

\section{Comparison in Supersymmetric Models}

Now we work on specific supersymmetric models. In trying to find scenarios where the LHC can probe better the new CP violating physics compared to eEDM measurements, we will work on models in which the eEDM is two-loop suppressed while TBV is only one-loop suppressed. As the simplest possibility we study the split supersymmetry limit where all scalars except SM-like neutral Higgs are heavy and decoupled~\cite{ArkaniHamed:2004fb,Giudice:2004tc,ArkaniHamed:2004yi,Wells:2003tf}. Another possibility is to take only first two generations of sleptons and squarks to be heavy, allow CP violating couplings in the trilinear scalar vertices of the third generation, which induces  radiative breaking of CP invariance in the Higgs sector. The mixing of CP even and CP odd eigenstates in the Higgs sector gives opportunity to colliders to discover these new sources of CP violation.

\subsection{Physical CP-phases and Triple Boson Vertices}\label{sec:CPphase}

One can see the relevance of TBVs in supersymmetric models in a more useful way using the physical CP-phases. Using $R$  and Peccei-Quinn ($PQ$) symmetries, it is shown that in any phase conventions there are two sets of physical CP-phases in the universality ansatz~\cite{Dimopoulos:1995kn,Dugan:1984qf}: arg($A \mu b^*$), arg($M_{1,2,3} \mu b^*$), where parameters are the usual soft supersymmetry parameters and higgsino mass $\mu$. The arg($M_iM_j^*$) are also allowed by the same argument. Since we impose GUT-like relations on gaugino masses these phases are not relevant to consider. As low-energy effective operators composed of SM Dirac fermions and vector fields are neutral under $R$ and $PQ$ symmetries, we can argue that low-energy physical observables should depend on the above combinations, which are the only $R$ and $PQ$ invariants. Indeed, both $R$ and $PQ$ are needed and enough for us to do that because all complex soft phases are charged under at least one of them. This argument does not restrict soft squark/slepton masses.

Since the $b$ term appears only in the Higgs sector, CP violation in the soft supersymmetry breaking sector can be transferred to low-energy effective operators consisting of Higgs bosons at one-loop order. Possible CP violating interactions with one SM Higgs field are Higgs-vector-vector, Higgs-fermion-fermion and Higgs-scalar-scalar couplings. Higgs-Higgs-vector coupling is usually related to the Higgs-vector-vector via the  underlying theory. As sizable tree-level processes at LHC involve at most two Higgs bosons, the scalar quartic coupling is not relevant.

As discussed in section \ref{sec:tbv}, there are also $SU(2) \times U(1)$ invariant dimension-six operators composed of Higgs bosons and vectors. After the Higgs bosons get vacuum expectation values, these operators can induce effective triple gauge couplings $WWV$, where $V$ is a neutral vector boson. Thus CP violating TBVs are not only relevant but also can indeed be generated at one-loop order in supersymmetric models. It is also clear that CP violating TBVs and EDMs depend on the same CP phases.

\subsection{Supersymmetry with Heavy Sfermions}

The split sfermion/ino limit of supersymmetry (split supersymmetry) does not naturally induce large EDMs. In this limit, charginos and neutralinos are not decoupled, and they carry CP phases in the soft supersymmetry breaking sectors. These ino sectors couple to SM fermions at tree-level only via ino-fermion-sfermion couplings which  lead to suppressed amplitudes in split supersymmetry due to the heavy sfermions. So CP violation in the SM fermion sector, e.g. EDM, are induced beginning at two-loop order. Recent studies have shown that the electron EDM turns out to be generically smaller than or around the current limit in most of parameter space even with maximum CP-phases~\cite{Giudice:2005rz,Chang:2005ac}.

To compute the effects, the input parameters are $\mu, \tilde{M}_{1,2}$ and their phases, $\tan \beta$ and SM-like neutral Higgs mass $M_h$. The sign of $\mu$ is not relevant as it just shifts the CP-phase by $\pi$. Since we are interested in electron electric dipole moments, the gluino mass $M_3$ is not relevant. Once we assume GUT-like relation between gaugino masses, only one CP-phase $\arg(\tilde{M} \mu b^*)$ is physical. The phase of $b$ is related to the relative phase of $H_u$ and $H_d$ via the  minimization condition of the Higgs potential; $b/(v_u v_d)$ is real at tree-level. We will work in the basis in which $b$ is real, then the two Higgs bosons have opposite phases. $U(1)_Y$ rotations of $H_u$ and $H_d$ can remove this relative phase, and the Higgs boson vevs are real in the same basis~\cite{Martin:1997ns}. The only physical combination of CP-phases remaining is $\arg(\tilde{M} \mu)$. It is clear that these CP-phases reside in the chargino and neutralino sectors.

\begin{figure}
\centering
\includegraphics[angle=0,width=0.95\textwidth]{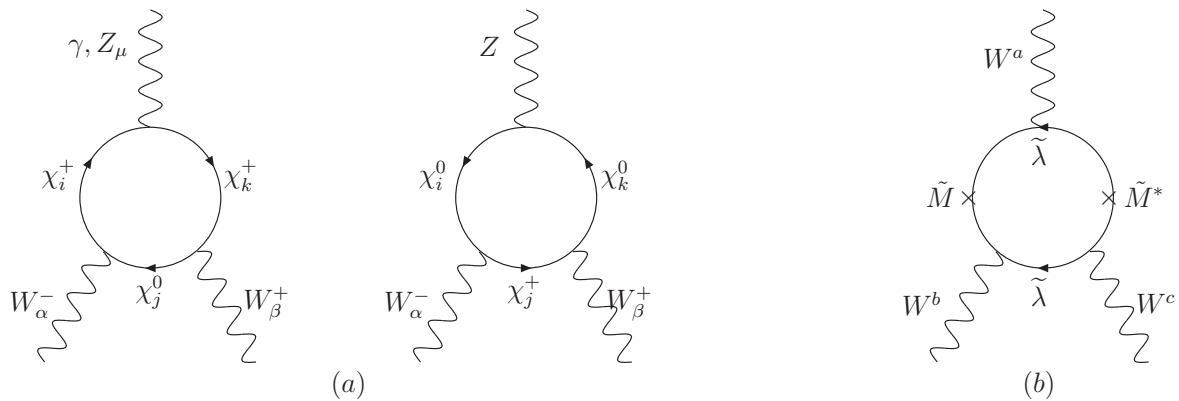}
\caption{(a) CP-odd TBV diagrams mediated by charginos $\chi^+_i\, (i=1,2)$ and neutralinos $\chi^0_i\,  (i=1,4)$. Similar diagrams generating $hVV$ couplings can also be drawn. (b) Diagram that is responsible for the $\tilde\lambda_V$ coupling is shown in terms of current eigenstates. Gaugino $\tilde\lambda$ is running in the loop, and its complex soft mass insertion is denoted as a cross. A similar diagram in which higgsinos are running with mass $\mu$ insertions can also be drawn.}\label{TBV:split}
\end{figure}

EDMs in split supersymmetry have been computed in previous works~\cite{Giudice:2005rz,Chang:2005ac,Feng:2008nm}. We also compute the effective CP-odd TBVs generated by diagrams shown in Fig.~\ref{TBV:split}(a), and apply them to the eEDM and collider observables. We give supporting analytic results in the Appendix.

One interesting result to notice is that the $\tilde\lambda_V$ coupling is not generated at one loop. It is simply because this coupling is generated by a dimension-six operator that does not involve Higgs fields as discussed in section~\ref{sec:tbv}, whereas the physical CP phase depends on the $b$ term. We can see this more explicitly in terms of current eigenstates depicted in Fig.~\ref{TBV:split}(b). As Higgs-higgsino-gaugino coupling couples gaugino and higgsino, either a gaugino or a higgsino runs in the loop without Higgs. Then $\mu$, as an interaction between $\tilde{H}_u$ and $\tilde{H}_d$, and $\tilde{M}_{1,2}$ cannot appear together, and hence no CP-phase. (Recall that the physical phases are $\mu M_{1,2}$ in our basis.) Indeed, the diagram with only gaugino (or higgsino) is proportional to $|\tilde{M}|^2$ (or $|\mu|^2$) because of the charge flow direction as shown in the figure. These are real, i.e.,  no CP violation.

Both CP-violating TBVs and eEDM are approximately proportional to $\sin 2\beta$ by essentially the same reason. To see this it is again easiest to think in terms of current eigenstates. Relevant diagrams are then Fig.~\ref{TBV:split}(b) with the $W$ boson on top replaced by a neutral gauge boson, and with mass insertions replaced by external $H_u, H_d$ legs and their vevs. Note that we need one $H_u$ and one $H_d$ in order to insert both $\tilde{M}$ and $\mu$. As we take neutral Higgs fields other than SM-like  Higgs boson to be very heavy, we obtain a simple relation between Higgs mixing angle $\alpha$ and vev ratio $\beta$: $\tan \alpha = \tan \beta$ at leading order. Therefore, each vev of $H_u$ and $H_d$ carries $\sin \beta$ and $\cos \beta$ respectively, hence $\sin 2\beta$ overall. eEDM is generated by inserting these effective interactions in Fig.~\ref{EDM:one-two}, thus having the same $\sin 2\beta$ dependence.

We now look at some numerical results for this scenario.
eEDM and CP violating TBVs depend on input parameters quite similarly. $\tan \beta$ dependence cancels when we study the relative importance of eEDM and collider observables as we saw above. Heavy $M_h$ can suppress the eEDM since Higgs boson mediated two-loop eEDM dominates numerically in this scenario, while TBVs are independent of $M_h$. However, due to the narrow consistent Higgs mass range $115 \GeV \leq M_h \lesssim 150 \GeV$ of the light SM-like Higgs boson in supersymmetry, this suppression is not very significant. $M_1$ dependence is weak since the bino does not couple to gauge bosons at tree-level. Dependence on the remaining gaugino/higgsino mass parameters can be different because the eEDM is two-loop while TBVs are one-loop physics.
\begin{figure}
\centering
\includegraphics[angle=0,width=0.45\textwidth]{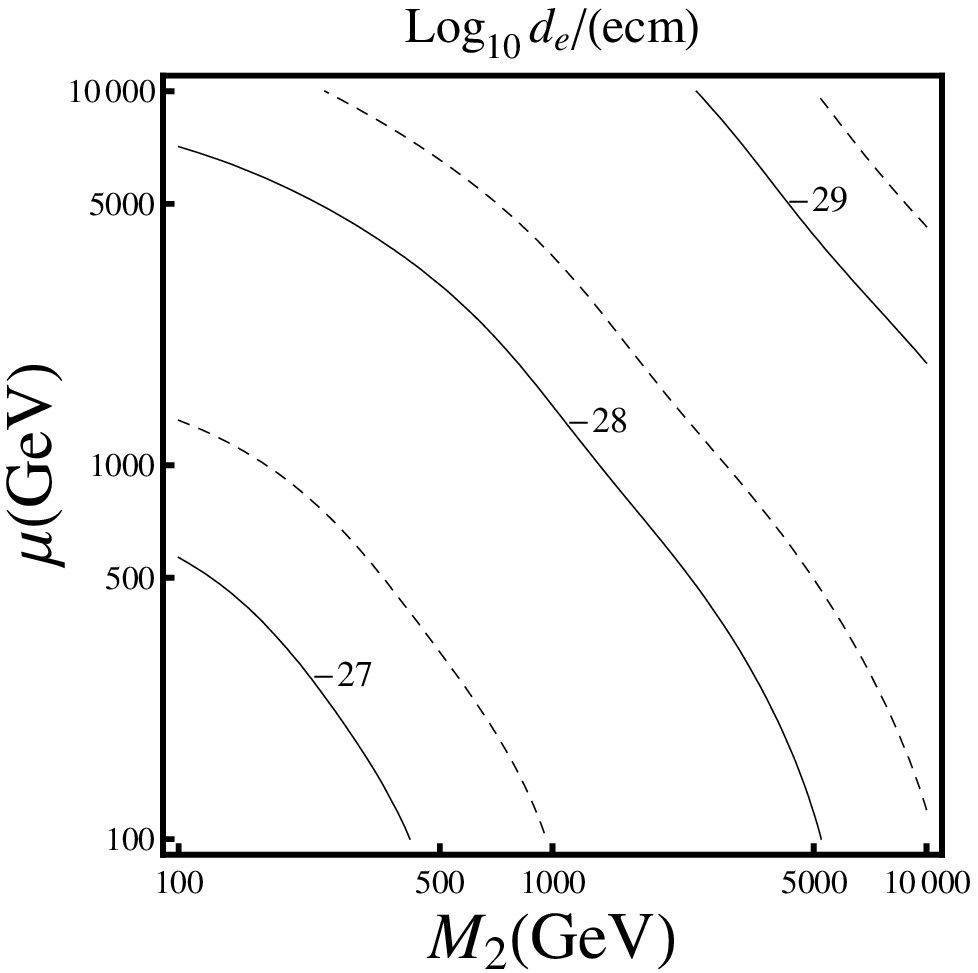}
\includegraphics[angle=0,width=0.45\textwidth]{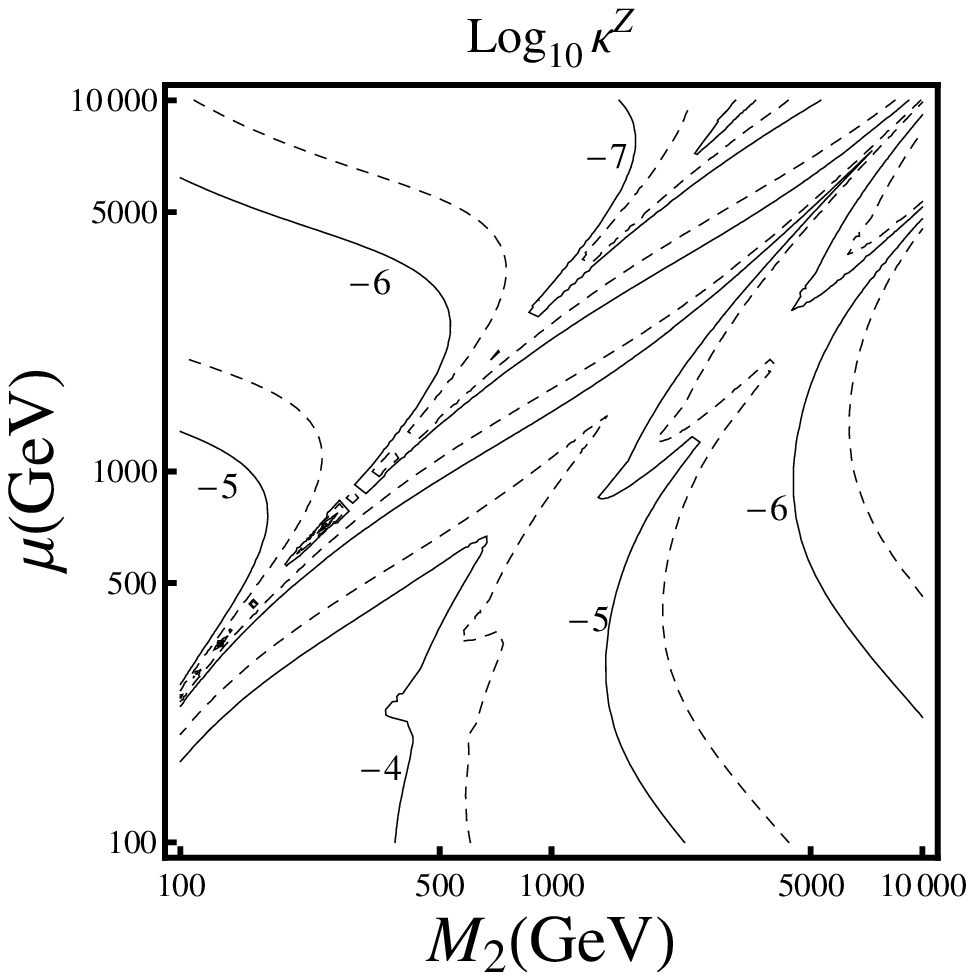}
\includegraphics[angle=0,width=0.45\textwidth]{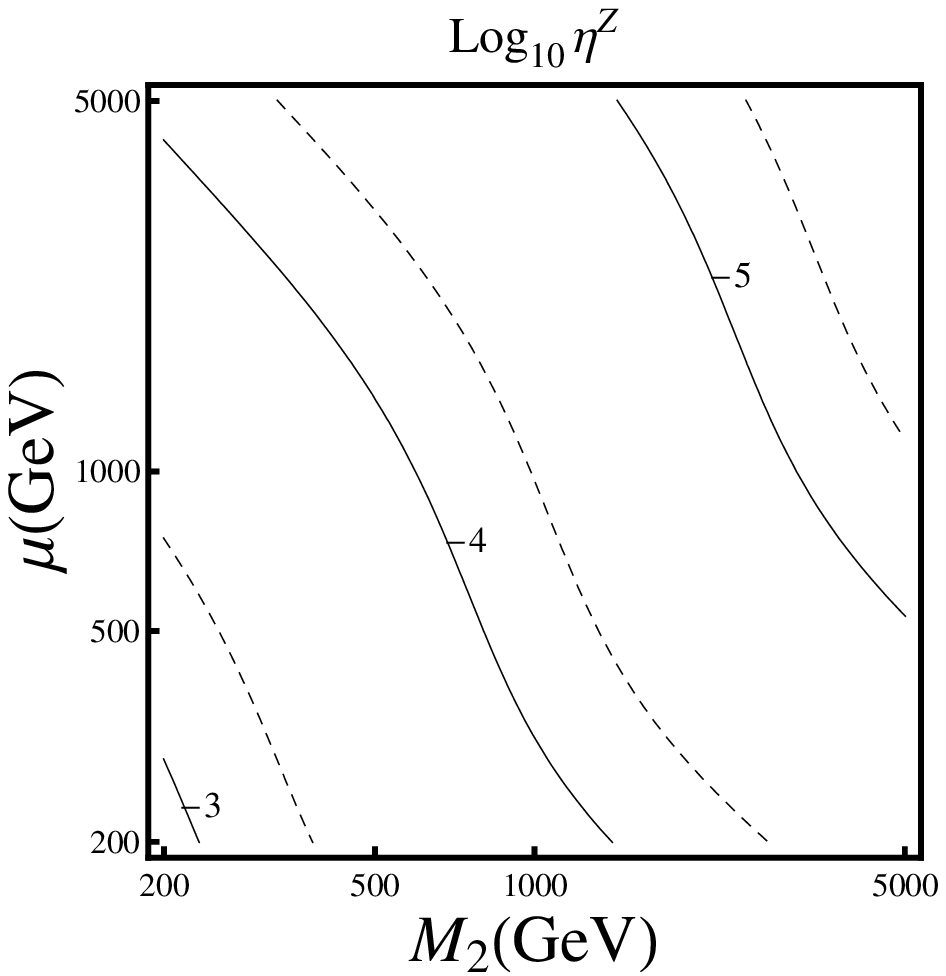}
\caption{Contour plots of eEDM, triple vector coupling $\tilde\kappa^Z$ and the Higgs coupling $\tilde\eta^Z$ to the $Z$ boson in the $M_2 - \mu$ plane. $\log_{10}$ values are written on the solid contour lines. To facilitate rescaling by the reader, contours are made for $\tan\beta=1$ with maximum CP phases. $M_h =120\GeV$ is used. Abrupt changes of $\tilde\kappa^Z$ in the diagonal region are partially due to a change of sign.}
\label{contour:split} \end{figure}

We choose to draw plots in $M_2 - \mu$ plane. In Fig.~\ref{contour:split}, we show eEDM and CP violating TBVs in this plane. We set $\tan\beta=1$ which is not allowed because this small $\tan\beta$ induces too large Yukawa coupling, but one can extrapolate the results linearly with $\sin2\beta$ as discussed above. In almost all of the parameter space, the current eEDM limit and the expected collider observable are not sensitive enough to probe CP violations in split supersymmetry even with maximum CP-phases.

Then the next question is if there exists parameter space in which eEDM is well below the future sensitivity while TBVs are around the future reach. The answer is (almost) no. In order to see this we scatter input parameters randomly within the following range:
\beq
100 \GeV \leq M_{1,2}, \mu \leq 1000\GeV , \quad 115\GeV \leq M_h \leq 180\GeV, \quad 2 \leq \tan \beta \leq 50.
\label{range:split} \eeq
If $M_2, \mu$ are a few $\TeV$, then both eEDM and collider observable are well below the current sensitivities as can be seen in \Fig{contour:split}, so we now focus on the sub-TeV gaugino/higgsinos. In addition, as stated earlier, we identify the future eEDM sensitivity to be $10^{-29} \ecm$ for reference~\cite{Kawall:2003ga,Hudson:2002,Liu:edm}.

\begin{figure}
\centering
\includegraphics[angle=0,width=0.47\textwidth]{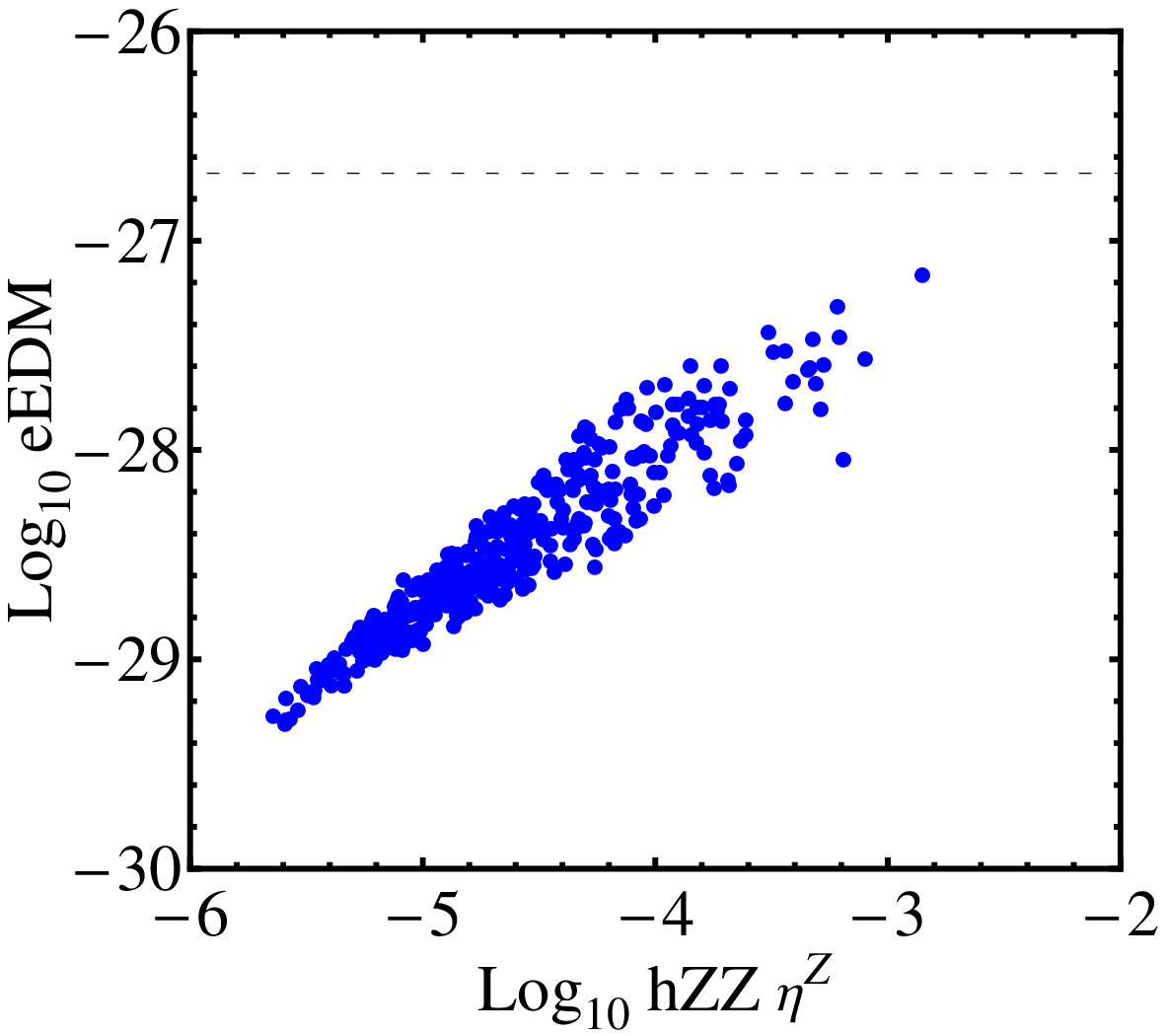}
\includegraphics[angle=0,width=0.47\textwidth]{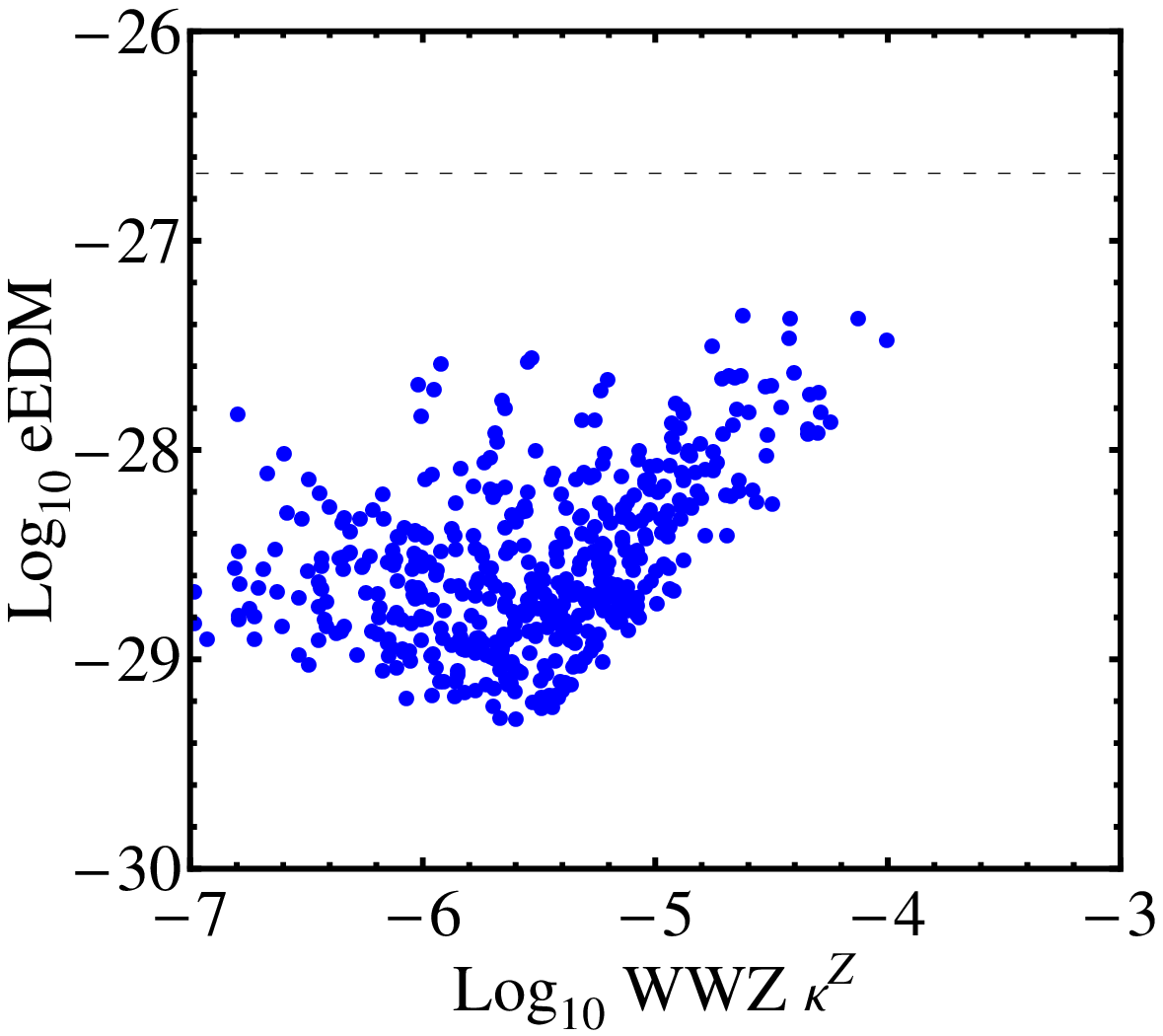}
\caption{CP violating Higgs (left) coupling $\tilde\eta^Z$ and triple vector (right) couplings $\tilde\kappa^Z$ to the $Z$ boson are plotted against eEDM in split supersymmetry. Input parameters are randomly scattered within the range Eq.(\ref{range:split}). The dashed horizontal line represents the current experimental eEDM bound $d_e<2.14\times 10^{-27}\ecm$.}
\label{EDMTBV:split} \end{figure}

In Fig.~\ref{EDMTBV:split}, we see that the eEDM and TBVs are closely related so that there is a  narrow allowed region of eEDM for each specific TBV value, and vice versa. The Higgs boson coupling shows stronger correlation with the eEDM due to dominance of the Higgs-mediated eEDM over $WW$-mediated eEDM. This correlation is what we expected based on the observation that any CP-violating TBV can induce an eEDM discussed in section~\ref{sec:edm}.

For $d_e < 10^{-29} \ecm$, which is just below the reference point of future eEDM measurement sensitivity, CP violating TBV values correspond to 
$\tilde\eta^Z,\, \tilde\kappa^Z \lesssim  8\times 10^{-6}$. Although it remains to be seen how well dedicated LHC experiments can do, if other CP violating observable expectations are a rough guide it is unlikely that these couplings can be probed at the one part per mil level at the LHC. If LHC fails to reach that very high sensitivity, the proposed eEDM sensitivity of $\sim 10^{-29}\ecm$ would be a more powerful probe of CP violation from new physics.

\begin{figure}
\centering
\includegraphics[angle=0,width=0.5\textwidth]{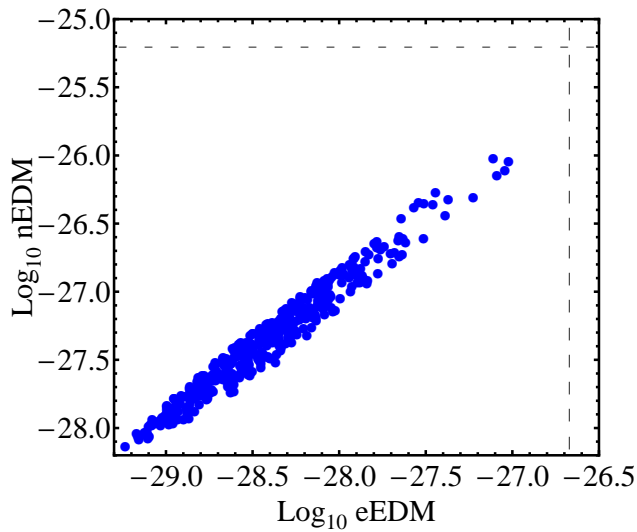}
\caption{Scatter plot of electron EDM and neutron EDM in split supersymmetry. Input parameters are scattered within the range \Eq{range:split}. Dashed lines represent the current experimental sensitivities.}
\label{eEDMnEDM:split} \end{figure}

The neutron EDM is also precisely measured with the current sensitivity~\cite{nEDM:sensitivity} $d_n < 6.3 \times 10^{-26} \ecm$, and can be improved in the future. This can be a competitor to the eEDM measurement depending on the future improvement and the theory prediction of the neutron EDM. In split supersymmetry, we compare computed eEDM and neutron EDM in \Fig{eEDMnEDM:split}. The neutron EDM is generated by constituent quark EDMs induced by the same types of diagrams generating eEDM because heavy squarks suppress chromo-electric dipole and three-gluon Weinberg operators. Thus, the neutron EDM depends on the same CP phases as eEDM does, and is closely related to eEDM as can be seen in \Fig{eEDMnEDM:split}~\cite{Abel:2005er}.

\subsection{MSSM with Radiative Breaking of Higgs Sector CP Invariance}

We relax the split limit but keep the first two generations of squarks and sleptons to be very heavy to avoid large FCNC and one-loop induced EDM~\cite{Abel:2001vy, moreminimal}. There are now not only additional physical CP phases, but the CP invariance of the Higgs sector can be radiatively broken so that there might be less correlation between eEDM and CP violating TBVs.

In the low-energy effective theory, we have two more neutral Higgs bosons, charged Higgs bosons and a third generation of squarks and sleptons in addition to split limit field contents. As a trilinear $A$-term interaction with stop (in large extent with sbottom) becomes relevant, the physical CP phases $\arg(A \mu b^*)$ cannot be ignored. These CP phases induce CP violation in the two-point Green's function through squark and quark loops, and mix CP-even and odd Higgs eigenstates~\cite{Pilaftsis:1998pe,Pilaftsis:1998dd}.  Because of these loop-induced interactions, we call this  ``radiative breaking" of CP invariance in the Higgs sector. One consequence of this important to us is that the pseudoscalar Higgs interactions with fermions generate CP violating TBVs at one-loop order. Of course, TBVs induce EDMs and the tension between them still exists.


However, the radiative breaking of CP invariance can enhance the CP violating collider observables. The neutral Higgs mass mixing matrix $O$ is defined as
\beq
\bmat H_d \\ H_u \\ A \emat = O \bmat H_1\\H_2\\H_3 \emat
\eeq
with $M_{H_1} < M_{H_2} < M_{H_3}$. $H_1$ $(H_2)$ becomes the light (heavy) CP-even Higgs in the absence of CP-violation. The scalar-pseudoscalar transitions $O_{Ai}$ induce pseudoscalar couplings between Higgs boson $H_i$ and quarks. Therefore, CP violating $H_iVV$ couplings generated by quarks are proportional to $O_{Ai}$ as explicitly shown in Eq.(\ref{quarkcont}). Other mixing elements modify CP conserving $H_iVV$ couplings. The ratio of the CP-even $H_iVV$ coupling in this scenario to the SM $H_iVV$ coupling is written as ~\cite{Pilaftsis:2000au}
\beq
g_{H_iVV} \, \equiv \,  c_\beta O_{H_di}\, +\, s_\beta O_{H_ui}.
\eeq
As scalar-pseudoscalar mixing $O_{Ai}$ increases, $g_{H_iVV}$ decreases because the mixing matrix is normalized. Thus the ratio of the CP-odd Higgs couplings to CP-even Higgs couplings can be relatively enhanced; i.e., the collider observable can be larger than what is expected in the case of no CP even-odd mixing. It is interesting to study if this enhancement can win over the limited amount of CP-violation allowed due to the eEDM bound.

CP even-odd mixing is large between two heavy neutral Higgs states $H_2$ and $H_3$ while the lightest Higgs $H_1$ remains mostly CP-even~\cite{Demir:1999hj}. So the enhancement is larger for $H_{2,3}$ couplings than for $H_1$. Meanwhile, as $g_{H_iVV}$ decreases we have to worry about a decrease of the cross section of diboson production mediated by Higgs bosons in Fig.~\ref{diboson:LHC}. We focus on the $gg \rightarrow H_i \rightarrow ZZ \rightarrow 4l$ diboson production channel for Higgs couplings collider observable as mentioned in section~\ref{sec:coll}. In order to use a collider observable, we need to be able to obtain at least a certain number of asymmetric events at the LHC. From this point of view, $H_1$ is a more important contributor than heavy Higgs bosons because $H_1$ is lighter and has larger couplings to the SM states. For example, the heavy Higgs $H_3$, which becomes a CP-odd eigenstate in the limit of no CP-violation, usually has very small CP-even $H_3VV$ couplings so there is little hope to measure them.

We now discuss the computation and numerical results for this scenario.
CP violating couplings are generated by Barr-Zee type diagrams in analogy to~\Fig{TBV:split}(a). In addition to gauginos and higgsinos, the third generation squarks and quarks can run in the loop~\cite{Pilaftsis:2002fe}. However, complex squark mixing angles cancel between adjacent vertices so squarks contribute to TBVs only at higher order. Top and bottom quarks can now generate CP violating Higgs couplings through tree-level pseudoscalar coupling. Meanwhile, the triple vector couplings are not affected by quarks, and not very different from the split supersymmetry case. Thus we focus on Higgs couplings in this section. Analytic results of quark and -ino contributions are shown in Appendix B. The complete set of two-loop induced EDMs in supersymmetry are computed in~\cite{Pilaftsis:2002fe,Li:2008kz,West:1994} and references therein.

$\widetilde{\lambda}_V$ couplings are still not generated at one-loop order in our analysis. Physical CP phases $\arg(A\mu b^*)$ and $\arg(M\mu b^*)$ depend on the $b$ term, so the same argument in split supersymmetry  case that forbade $\tilde\lambda_V$ applies here as well. We can take another linear combination $\arg(AM^*)$, which appears at two-loop order, as squark and gaugino couple through a triple vertex with a quark. This $SU(2)$ analogy of the three-gluon Weinberg operator has little effect on the eEDM, as discussed in section~\ref{sec:edm}.

We assume the universality and flavor-diagonality of soft masses and the trilinear coupling $A$-term for simplicity. The input parameters are then
\beq
M_{1,2},\quad \mu,\quad \tan \beta,\quad M_{H^\pm},\nonumber\\ \eeq
\beq
A=A_t=A_b=A_\tau,\quad M_{SUSY}=M_{Q_3}=M_t=M_b=M_{L_3}=M_\tau
\eeq
and soft CP-phases. As heavy Higgs bosons are not decoupled, the Higgs boson mixing angle $\alpha$ is not trivially related to vev ratio $\beta$, i.e. $\tan\alpha \neq \tan\beta$. The Higgs boson mixing angle now depends on various input parameters. Then the previous argument about $\sin2\beta$ dependence in split supersymmetry does not apply here. Indeed, several authors have shown that the eEDM increases overall with $\tan\beta$~\cite{Pilaftsis:2002fe}. Here, $\tan\beta$ rather plays the role of determining the amount of enhancement through $g_{H_iVV}$ and couplings with fermions which can also be seen in \Eq{quarkcont}.

We have modified the CPsuperH 2.0 program~\cite{CPsuperH} for numerical study. We scattered input parameters within the range
\beq
300\GeV \leq A,M_{SUSY} \leq 2000\GeV, \nonumber\\ \eeq
\beq
130\GeV \leq M_{H^\pm} \leq 250 \GeV ,\quad 150\GeV \leq M_{1,2} \mu \leq 1000\GeV, \quad 2\leq t_\beta \leq 50.
\eeq
We also consider the following consistency condition
\beq
M_{H_1} \geq 115\GeV .
\label{condition:MSSM} \eeq

\begin{figure}
\centering
\includegraphics[angle=0,width=0.5\textwidth]{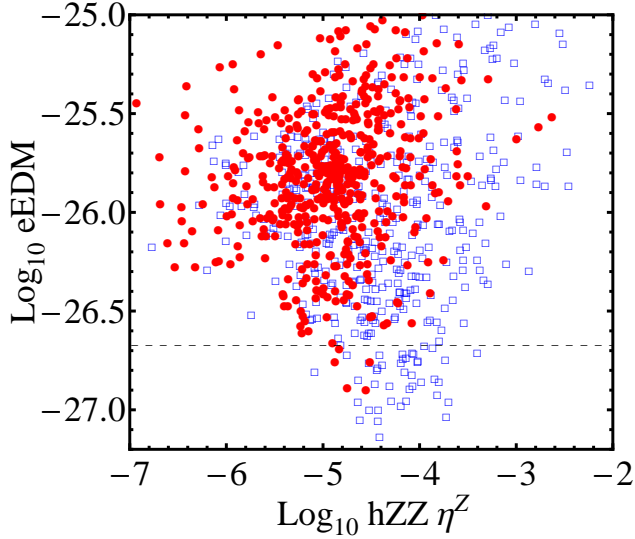}
\caption{Light Higgs $H_1$ CP violating coupling $\tilde\eta^Z$ to the $Z$ boson is plotted against eEDM with light third generation squarks. Blue $\Box$ are excluded by $M_h \geq 115\GeV$ while red $\bullet$ satisfy this condition. Dashed line represents current eEDM bound $d_e < 2.14\times 10^{-27}\ecm$. Expected sensitivity on Higgs coupling is too large to be shown. This plot is generated with maximum CP-violating phases, and all points move downward as the phase angles decrease.}
\label{hzz-edm-2}
\end{figure}

The light Higgs boson $H_1$ coupling to the $Z$ boson (see \Eq{CPodd:WWV} for definition) versus the computed eEDM is shown in \Fig{hzz-edm-2}. Sample points which satisfy consistency condition in Eq.(\ref{condition:MSSM}) are represented as red circles. The eEDM measurement alone eliminates most of the sample points and restricts the Higgs coupling to be well below the experimental sensitivity $\sim O(0.1)$. Actually, in  most of parameter space consistent with condition Eq.(\ref{condition:MSSM}) and eEDM bound, $g_{H_1VV} \sim 1$ and $O_{A1} \lesssim O(0.01)$. Thus, enhancement is too small to overcome the eEDM constraint. Large CP violation needed to obtain large $O_{A1}$ and small $g_{H_1VV}$ is still prohibited by the eEDM constraint.

For heavy Higgs boson $H_2$, it also turned out to be very pessimistic for collider signatures of CP violation. The required cross-section just to discover the $H_2$ Higgs boson itself almost eliminates the possibility for us to measure a $H_2VV$ CP violating couplings. This cross section is at best a few hundred $ab$ for $M_{H_2} \sim 170 \GeV$.

In this particular limit, the neutron EDM is usually predicted to be about two orders of magnitude larger than the eEDM, and hence is a stronger constraint on new physics~\cite{Abel:2005er}. This is mainly because the large $A_t$ coupling generates a three-gluon Weinberg operator that dominantly contributes to the neutron EDM while the stop contribution to eEDM is subdominant. In any case, large CP violations generating CP violating TBVs eventually induce EDM, which is generally more constraining than CP-violating collider physics observables.


In concluding this section, we mention the previous work of Babu et al.~\cite{Babu:1998bf}, which had similar goals of comparison as this work. We briefly discuss that paper since it strengthens our conclusion. They found that one-loop lepton EDM mediated by slepton and gaugino/higgsino puts severe constraint on one-loop generated CP violating Higgs-lepton-lepton couplings in the MSSM. So they sought other places where CP violating Higgs boson couplings may be  enhanced while the lepton EDM is relatively not. They noted that there is a tree-level CP even-odd mixing in the Higgs sector of the NMSSM. Since this CP violating coupling is not loop suppressed and has different dependence on input parameters than the one-loop lepton EDM, they suggested that this would be a good place to observe large CP violating Higgs couplings. However, once Higgs-mediated two-loop EDM contributions are considered this conclusion must be modified. The large CP violating Higgs boson couplings induce a two-loop EDM regardless of the origin of such CP violating couplings, which constrains the size of these Higgs boson couplings quite severely.

\section{Conclusions}

Our basic conclusion, which is supported by detailed investigations of various candidate theories that had a chance to contravene it, is this: Whatever the origin may be of CP violating triple boson vertices, they induce EDMs, and although the physics that induces EDMs is ``one loop down" compared to collider CP asymmetries, the EDM experiments are sufficiently precise that they overcome the loop factor and are generally more powerful probes. We expect this conclusion to strengthen into the foreseeable future as EDM experiments become more sensitive.

\section*{Acknowledgements}
We  thank R. Akhoury, J. Kumar, A. Leanhardt, A. Pierce, A. Rajaraman and J. Shao for helpful discussions. This work is supported in part by the Department of Energy. S. Jung is also supported in part by Samsung Scholarship.

\section*{Appendix A - CP Violating Couplings}
In this appendix, we present our conventions and analytic results of CP violating TBVs.

\subsection*{A.1 Conventions}
Gaugino and higgsino masses are given as
\beq
-{\cal L} = \frac{1}{2} M_1 \widetilde{B} \widetilde{B} + \frac{1}{2} M_2 \widetilde{W}^a \widetilde{W}^a + \mu \widetilde{H_u} \epsilon \widetilde{H_d}.
\eeq
Chargino, neutralino mixing matrices $U,V,N$ satisfy
\beq
N^* M_{\chi^0} N^\dagger = M_{D^0} , \qquad U^* M_{\chi^+} V^\dagger = M_{D^+}
\eeq
where $M_{\chi^0}$ and $M_{\chi^+}$ are as in Ref.~\cite{Martin:1997ns}. The subscript D implies a diagonal matrix with positive elements.

The interaction Lagrangian of split supersymmetry in terms of mass eigenstates is
\bea
{\cal L} &=& g \overline{\chi_i^0} \gamma^\mu \left( C_{ij}^L P_L + C_{ij}^R P_R \right) \chi_j^+ W_\mu^+ + h.c. \nonumber\\ && +\frac{g}{c_W} \overline{\chi_i^+} \gamma^\mu \left( F_{ij}^L P_L + F_{ij}^R P_R \right) \chi_j^+ Z_\mu + \frac{g}{c_W} \overline{\chi^0_i} \gamma^\mu ( H^L_{ij} P_L + H^R_{ij} P_R ) \chi^0_j Z_\mu \nonumber\\
&& + \frac{g}{\sqrt{2}} \overline{\chi_i^+} \left( D_{ij}^L P_L + D_{ij}^R P_R \right) \chi_j^+ h \,+ \, \frac{g}{\sqrt{2}} \overline{\chi_i^0} \left( D^{\prime L}_{ij} P_L + D^{\prime R}_{ij} P_R \right) \chi_j^0 h \nonumber\\
&& -e Q_f \bar{f}\gamma^\mu f A_\mu + \frac{g m_f}{2 M_W} \overline{f} f h
\label{lagran} \eea
where $f$ is a fermion for which the EDM is calculated. C,D,F and H are give by
\bea
C^L_{ij} &=& N_{i2} V^*_{j1} - \frac{1}{\sqrt{2}}N_{i4}V^*_{j2} , \qquad C^R_{ij} = N^*_{i2} U_{j1} + \frac{1}{\sqrt{2}}N^*_{i3}U_{j2}  \nonumber\\
F^L_{ij} &=& -\delta_{ij} c_W^2 + \frac{1}{2} V_{i2}V^*_{j2}, \qquad F^R_{ij} = -\delta_{ij} c_W^2 + \frac{1}{2} U^*_{i2}U_{j2} \nonumber\\
H^L_{ij} &=& -\frac{1}{4} ( N^*_{i3} N_{j3} - N^*_{i4} N_{j4} ) ,\qquad H^R_{ij} = -(H^L_{ij})^* = -H^L_{ji} \nonumber\\
D^L_{ij} &=& s_\beta U^*_{i1} V^*_{j2} + c_\beta U^*_{i2} V^*_{j1} , \qquad D^R_{ij} = s_\beta V_{i2} U_{j1} + c_\beta V_{i1} U_{j2} = (D^{L \dagger})_{ij} \nonumber\\
D^{\prime L}_{ij} &=& \left( N^*_{j2}- t_WN^*_{j1} \right) \, \left( N^*_{i3} c_\beta -  N^*_{i4} s_\beta \right) + (i \leftrightarrow j) , \qquad D^{\prime R}_{ij} = (D^{\prime L}_{ij} )^*
\label{couplings} \eea
Here, index $3 (4)$ implies $H_d (H_u)$ following Ref.~\cite{Martin:1997ns}.
In the MSSM away from the split supersymmetry limit, Higgs boson couplings are modified as the relation $\tan\alpha = \tan\beta$ does not generally hold. For the lightest Higgs boson $H_1$, the couplings can be obtained by substituting $h\rightarrow H_1$ and $s_\beta (c_\beta) \rightarrow -c_\alpha (s_\alpha)$ where $s_\beta$ and
$c_\beta$ are explicitly listed in the above equations for $D^{L,R}_{ij}$.

CP and P-odd form factors are conventionally written as below~\cite{Hagiwara:1986vm} for incoming $V_\mu (q)$ (or $h(q)$) and outgoing $W^-_\alpha(p_1)$ and $W^+_\beta (p_2)$ (or $V_\mu(p_1)$ and $V_\nu (p_2)$)
\bea
\Gamma^{\mu\alpha\beta}_{\textrm{$WWV$}} &=& ig_{WWV} \left[ f_6^V(q) \, \epsilon^{\mu\alpha\beta\nu} q_\nu \,+  \, \frac{f_7^V(q)}{M_W^2}(p_1-p_2)^\mu \epsilon^{\alpha\beta\rho\sigma} q_\rho (p_1-p_2)_\sigma \right. \nonumber\\
&&  \hspace{2.7in} +\, if_4^V(q)(q^\alpha g^{\mu\beta} + q^\beta g^{\mu\alpha})  \Big] \nonumber\\
\Gamma_{H_iVV}^{\mu\nu} &=& gM_W \, \left[ g_{H_iWW} \left( S_i^W(q) (g_{\mu\nu} - \frac{2 p_{1\mu} p_{2\nu}}{M_W^2}) + \frac{P_i^W(q)}{M_W^2} \epsilon_{\mu\nu\alpha\beta} p_1^\alpha p_2^\beta \right) \right.\ \nonumber\\ && \left.\ \qquad \qquad +\, \frac{1}{2 c_W^2} g_{H_iZZ} \left( S_i^Z(q) (g_{\mu\nu} - \frac{2 p_{1\mu} p_{2\nu}}{M_W^2}) + \frac{P_i^Z(q)}{M_W^2} \epsilon_{\mu\nu\alpha\beta} p_1^\alpha p_2^\beta \right) \right]
\label{formfactor} \eea
where $f_6^V=\tilde\kappa_V - \tilde\lambda_V,\, f_7^V = -\frac{1}{2} \tilde\lambda_V,\, f_4^V=g_4^V,\, P_i^V=\tilde\eta_i^V$. $g_{WW\gamma}=-e,\, g_{WWZ}=-e\cot\theta_W$ and $g_{H_iVV}$ is the ratio of the CP-even $H_iVV$ coupling to the SM $H_iVV$ coupling. CP-even form factor $S_i^V$ and C-odd form factor $f_4^V$ are shown for reference. More information about these form factors and the effective Lagrangian \Eq{CPodd:WWV} can be found in~\cite{Hagiwara:1986vm}.

\subsection*{A.2 Triple Boson Vertices}
We represent triple gauge boson form factors first. These are generated via chargino/neutralino as shown in Fig.~\ref{TBV:split}. The effective couplings are obtained in the limit of the on-shell center-of-mass energy $s =q^2 \rightarrow M_{V,H_i}^2$. For reference we list all three types of CP violating $WWV$ couplings in terms of loop functions $a_i^{WWV}$.
\bea f_6^{Z++} &=& \frac{g^2}{16\pi^2c_W^2} \sum_{i,j,k}  \left[ m_i^+ m_k^+ \Im(C_{ji}^{R*} C_{jk}^R F_{ki}^L -L) a_1^{WWZ}  + 2m_i^+ m_j^0 \Im(C_{ji}^{L*}C_{jk}^R F_{ki}^R-L) a_2^{WWZ} \right. \nonumber\\ & & \left. + 2\Im (C_{ji}^{R*} C_{jk}^R F_{ki}^R -L) \left\{ M_W^2 a_5^{WWZ} + q^2 (a_4^{WWZ}/2 - a_6^{WWZ}) + 3a_8^{WWZ} \right\} \right]
\eea
\bea f_4^{Z++} &=& \frac{g^2}{16\pi^2c_W^2} \sum_{i,j,k}  \left[ m_i^+ m_k^+ \Im(C_{ji}^{R*} C_{jk}^R F_{ki}^L +L) a_1^{WWZ}  + 2m_i^+ m_j^0 \Im(C_{ji}^{L*}C_{jk}^R F_{ki}^R+L) a_3^{WWZ} \right. \nonumber\\ & & \left. + 2\Im (C_{ji}^{R*} C_{jk}^R F_{ki}^R +L) \left\{ M_W^2 a_5^{WWZ} - q^2 a_6^{WWZ} + 3a_8^{WWZ} \right\} \right]
\eea
\bea f_6^{Z00} &=& \frac{g^2}{16\pi^2c_W^2} \sum_{i,j,k}  \left[ m_i^0 m_k^0 \Im(C_{ij}^{R*} C_{kj}^R H_{ik}^L -L) a_1^{WWZ}  + 2m_j^+ m_k^0 \Im(C_{ij}^{L*}C_{jk}^R H_{ik}^L-L) a_2^{WWZ} \right. \nonumber\\ & & \left. + 2\Im (C_{ij}^{R*} C_{kj}^R H_{ik}^R -L) \left\{ M_W^2 a_5^{WWZ} + q^2 (a_4^{WWZ}/2 - a_6^{WWZ}) + 3a_8^{WWZ} \right\} \right]
\eea
\bea f_4^{Z00} &=& \frac{g^2}{16\pi^2c_W^2} \sum_{i,j,k}  \left[ - m_i^0 m_k^0 \Im(C_{ij}^{R*} C_{kj}^R H_{ik}^L +L) a_1^{WWZ}  + 2m_j^+ m_k^0 \Im(C_{ij}^{L*}C_{kj}^R H_{ik}^L+L) a_3^{WWZ} \right. \nonumber\\ & & \left. + 2\Im (C_{ij}^{R*} C_{kj}^R H_{ki}^R +L) \left\{ M_W^2 a_5^{WWZ} + q^2 a_6^{WWZ} - 3a_8^{WWZ} \right\} \right] \\
f_7^Z &=& 0 \\
f_6^\gamma &=& \frac{e^2}{8\pi^2}\sum_{i,j} \, m_i^+ m_j^0 \Im(C^{L*}_{ji}C^R_{ji} - L) \, a_2^{WW\gamma} \\
f_4^\gamma &=& f_7^\gamma \,= \,0
\eea
Subscript $^{++} (^{00})$ implies the contributions from the first (second) diagram in Fig.~\ref{TBV:split} where two charginos (neutralinos) are running in the loop. $L$ inside the Im part implies the same coupling combination with $L \leftrightarrow R$. $f_4^\gamma$ is zero because $WW\gamma$ form factors define the electric charge of the $W$ boson in the Coulomb limit while C-odd parts flip the electric charge.

The loop functions are given as (assuming light on-shell bosons)
\bea
a_i^{WWZ} &=& \int^1_0 dx \int^{1-x}_0 dy \, \frac{b_i}{(m_i^2-m_j^2)x + (m_k^2-m_j^2)y + m_j^2 - q^2xy} \qquad \textrm{for } i=1, \cdots , 7 \nonumber\\
a_8^{WWZ} &=& \int^1_0 dx \int^{1-x}_0 dy \, (y-x) \cdot \log \left( (m_i^2-m_j^2)x + (m_k^2-m_j^2)y + m_j^2 - q^2xy \right) \nonumber\\ &=& (m_k^2 -m_i^2) \, a_7^{WWZ}
\eea
where $q$ is incoming $Z$ boson momentum. $a_2^{WW\gamma}$ can be obtained by taking $m_k=m_i$ in $a_2^{WWZ}$. Coefficients $b_i$ are given as
\bea
b_1 &=& x-y, \quad b_2=y-x+1, \quad b_3 = x+y-1, \quad b_4= (y-x)(x+y-1) \nonumber\\
b_5 &=&(y-x)(x+y-1)^2, \qquad b_6 = (y-x)\,xy, \qquad b_7 = xy.
\eea
These results numerically match well with previous computations~\cite{Kitahara:1997}.

In a similar way, $H_iVV$ couplings are generated via chargino/neutralino and top/bottom quarks (not in split supersymmetry). Here we represent only CP-odd $hZZ$ and $hWW$ couplings as these are relevant for our numerical studies. These are given in terms of loop functions $c_i$.
\bea
P^Z_h \cdot g_{hVV} &=& \frac{\sqrt{2} \alpha M_W}{ \pi s_W^2} \sum_{i,j,k=1}^2 \left( m_i \Im( F^R_{ji} D^R_{ik} F^R_{kj} - L ) c_1(i,j,k)  + m_j \Im(F^R_{ji} D^L_{ik} F^L_{kj} -L ) c_2(i,j,k) \right)  \nonumber\\
&+&  \frac{\sqrt{2} \alpha M_W}{ \pi s_W^2} \sum_{i,j,k=1}^4 \left( m_i \Im( H^R_{ji} D^{\prime R}_{ik} H^R_{kj} - L ) c_1(i,j,k)  + m_j \Im(H^R_{ji} D^{\prime L}_{ik} H^L_{kj} -L ) c_2(i,j,k) \right) \nonumber\\
&+&  \frac{3 \alpha M_W}{\pi s_W^2} \sum_{f=t,b} m_f \left( \Im(F^R_Z D^R_f F^R_Z -L) c_3(f) + \Im(F^R_Z D^L_f F^L_Z -L) c_4(f) \right).
\eea
\bea
P^W_h \cdot g_{hVV} &=& \frac{\sqrt{2} \alpha M_W}{ \pi s_W^2} \sum_{i,j,k=1}^2 \left( m_i \Im( C^R_{ji} D^R_{ik} C^{*R}_{kj} - L ) c_1(i,j,k)  + m_j \Im(C^R_{ji} D^L_{ik} C^{*L}_{kj} -L ) c_2(i,j,k) \right)  \nonumber\\
&+&  \frac{\sqrt{2} \alpha M_W}{ \pi s_W^2} \sum_{i,j,k=1}^4 \left( m_i \Im( C^{*R}_{ji} D^{\prime R}_{ik} C^R_{kj} - L ) c_1(i,j,k)  + m_j \Im(C^{*R}_{ji} D^{\prime L}_{ik} C^L_{kj} -L ) c_2(i,j,k) \right) \nonumber\\
&+&  \frac{3 \alpha M_W}{\pi s_W^2} \sum_{f=t,b} m_f  \Im(-D_f^L) \,c_3(f).
\eea
where couplings with quarks are given as
\bea
F_Z^{L,R} = T^3_f - Q_f s_W^2 ,\qquad D^L_t = \frac{m_t}{M_W s_\beta} (O_{H_u i} + i O_{Ai} c_\beta )&,& \quad D^R_t = (D^L_t )^* \nonumber\\
D^L_b = \frac{m_b}{M_W c_\beta} (O_{H_d i} + i O_{Ai} s_\beta )&,& \quad D^R_b = (D^L_b )^*
\eea
In order to see the dependence on CP even-odd mixing better, we simplify the quark contributions in the third lines by approximately treating $s_W^2 \approx 0.25$. These quark contributions are given as
\bea
P^Z_h \cdot g_{hVV} &\cong& - \frac{ 3 \alpha O_{A1} }{\pi s_W^2} \left\{ \frac{m_t^2}{t_\beta} \left(\frac{10}{72} c_3(t) + \frac{8}{72} c_4(t) \right) + m_b^2 t_\beta \left( \frac{13}{72} c_3(b) + \frac{5}{72} c_4(b) \right) \right\} \,+ \cdots \nonumber\\
P^W_h \cdot g_{hVV} &\cong& - \frac{ 3 \alpha O_{A1} }{\pi s_W^2} \left\{ \frac{m_t^2}{t_\beta}\, c_3(t) \,+\, m_b^2 t_\beta \, c_3(b) \right\} \,+ \cdots
\label{quarkcont} \eea
We can see that the CP-odd $hZZ$ coupling is very sensitive to the CP even-odd mixing $O_{A1}$ and $t_\beta$. Quantum corrections to the CP-even couplings are ignored as they are much smaller than the tree-level values.

The loop functions are (assuming on-shell vector bosons)
\bea
c_1(i,j,k) &=& \int^1_0 dx \int^{1-x}_0 dy \, \frac{x+y}{(m_i^2-m_j^2)x + (m_k^2-m_j^2)y + m_j^2 + M_V^2(x+y)(x+y-1) -q^2xy} \nonumber\\
c_2(i,j,k) &=& \int^1_0 dx \int^{1-x}_0 dy \, \frac{x+y-1}{(m_i^2-m_j^2)x + (m_k^2-m_j^2)y + m_j^2 + M_V^2(x+y)(x+y-1) -q^2xy} \nonumber\\
c_3(f) &=& \int^1_0 dx \int^{1-x}_0 dy \, \frac{x+y}{m_f^2 + M_V^2(x+y)(x+y-1) -q^2xy} \nonumber\\
c_4(f) &=& \int^1_0 dx \int^{1-x}_0 dy \, \frac{x+y-1}{m_f^2 + M_V^2(x+y)(x+y-1) -q^2xy}.
\eea
where $q$ is Higgs momentum.

\begin{figure}
\centering
\includegraphics[angle=0,width=0.42\textwidth]{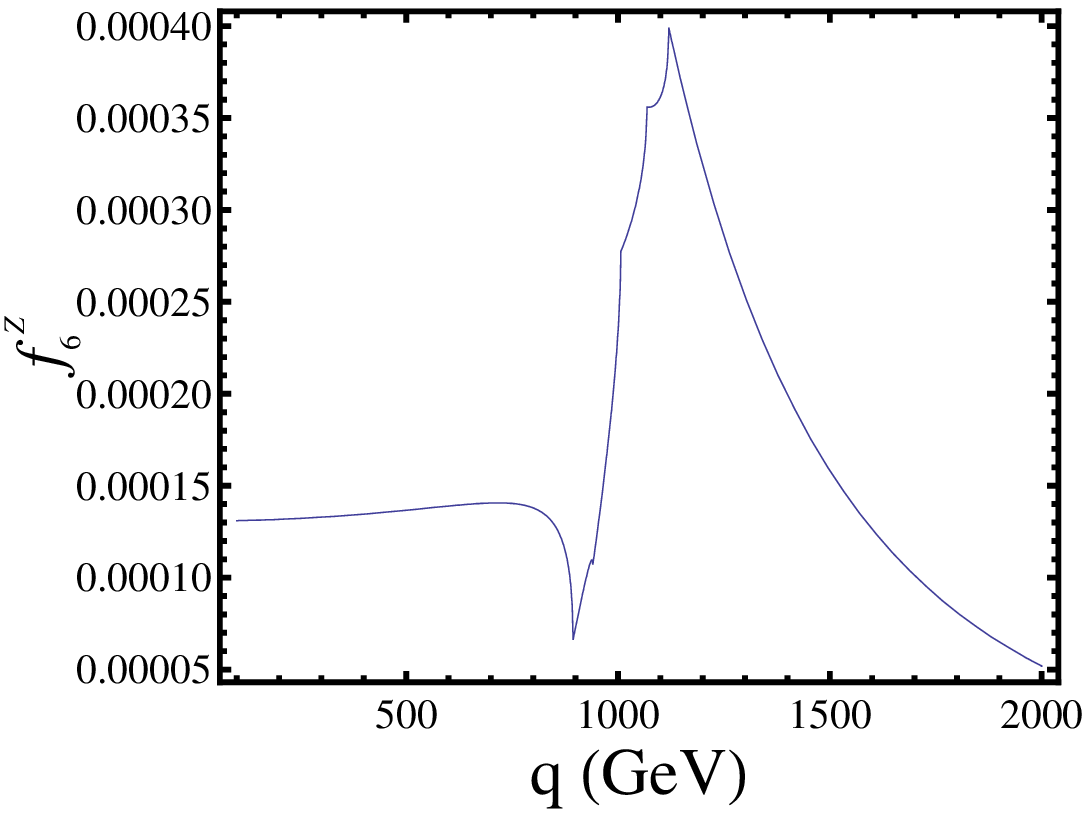}
\includegraphics[angle=0,width=0.42\textwidth]{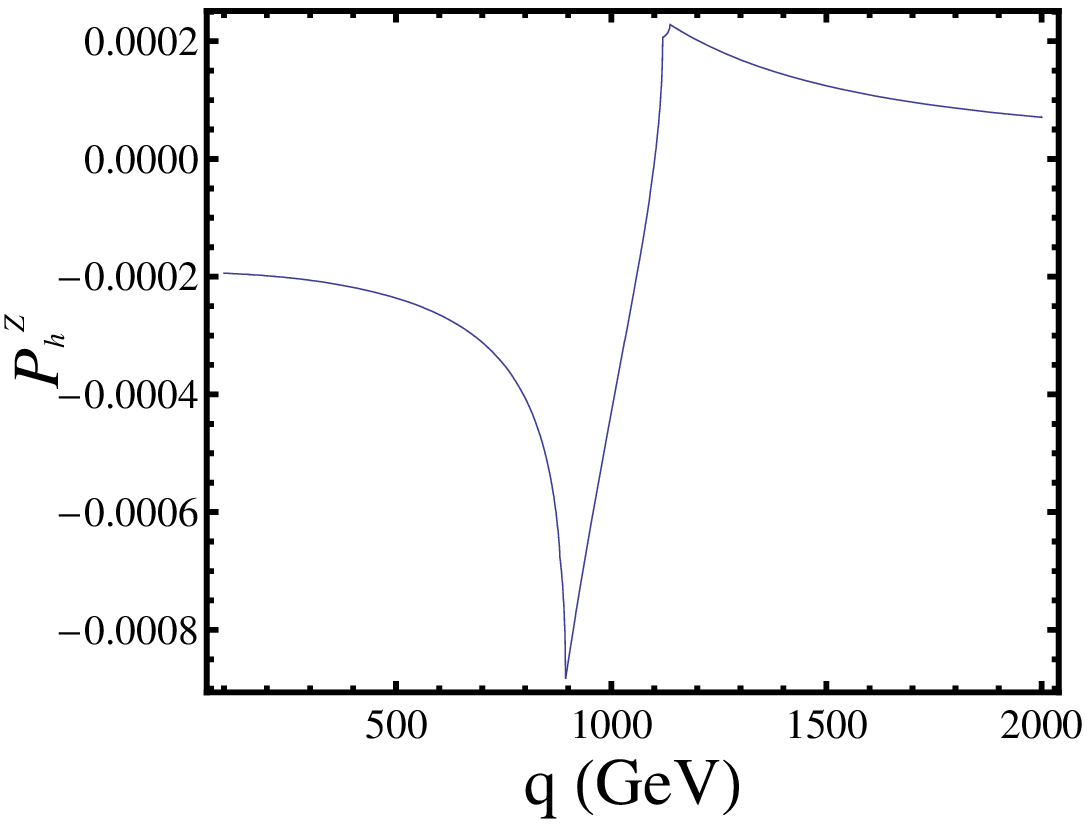}
\caption{Sample plots show the momentum dependence of form factors $f^Z_6$ (left) and $P^Z_h$ (right) in split supersymmetry. $q$ is $Z$ or Higgs momentum. $M_1=M_2=\mu=500 \GeV$ and $t_\beta =1$ are used.}
\label{momdep}
\end{figure}

\begin{figure}
\centering
\includegraphics[angle=0,width=0.48\textwidth]{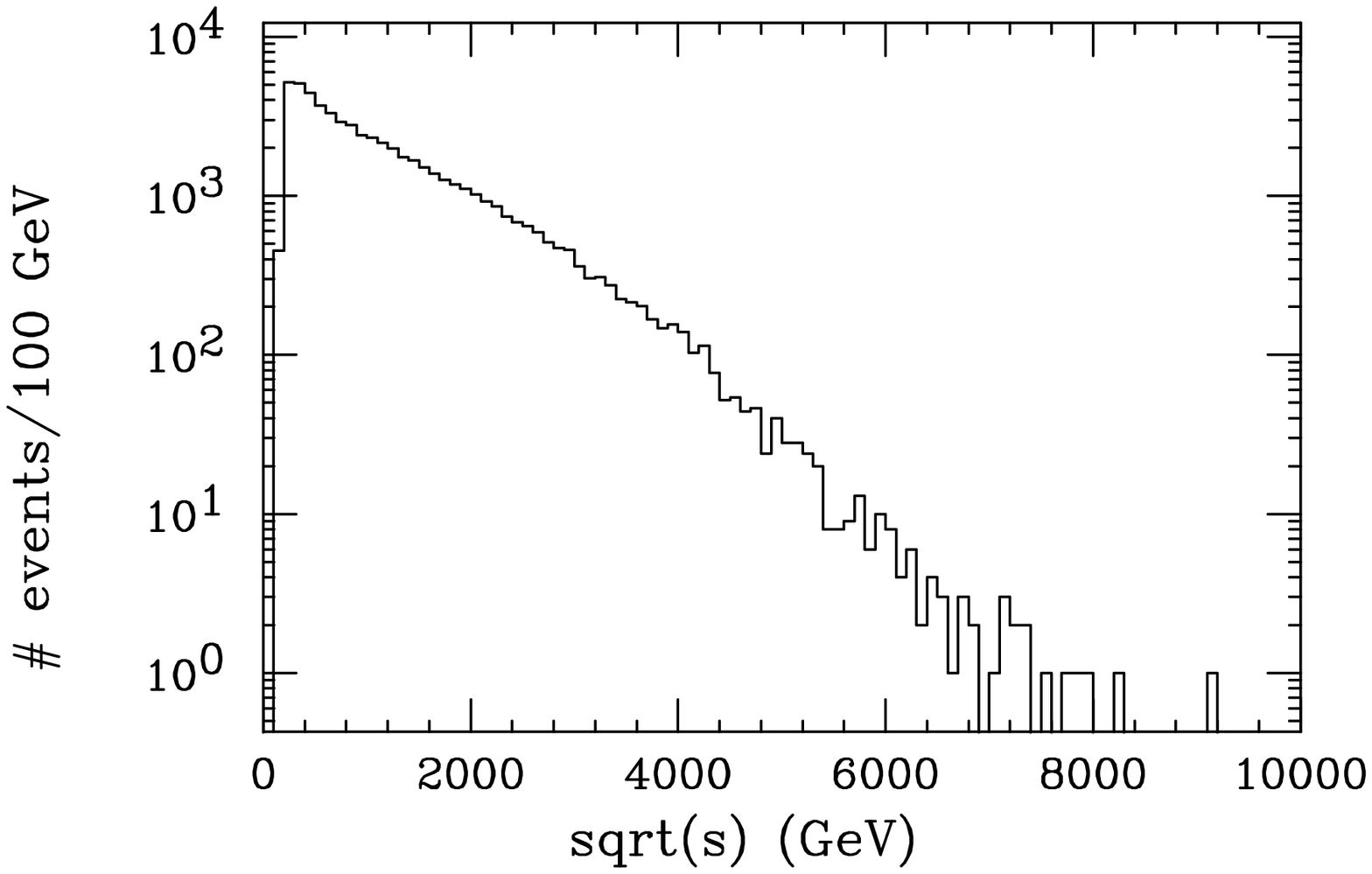}
\includegraphics[angle=0,width=0.48\textwidth]{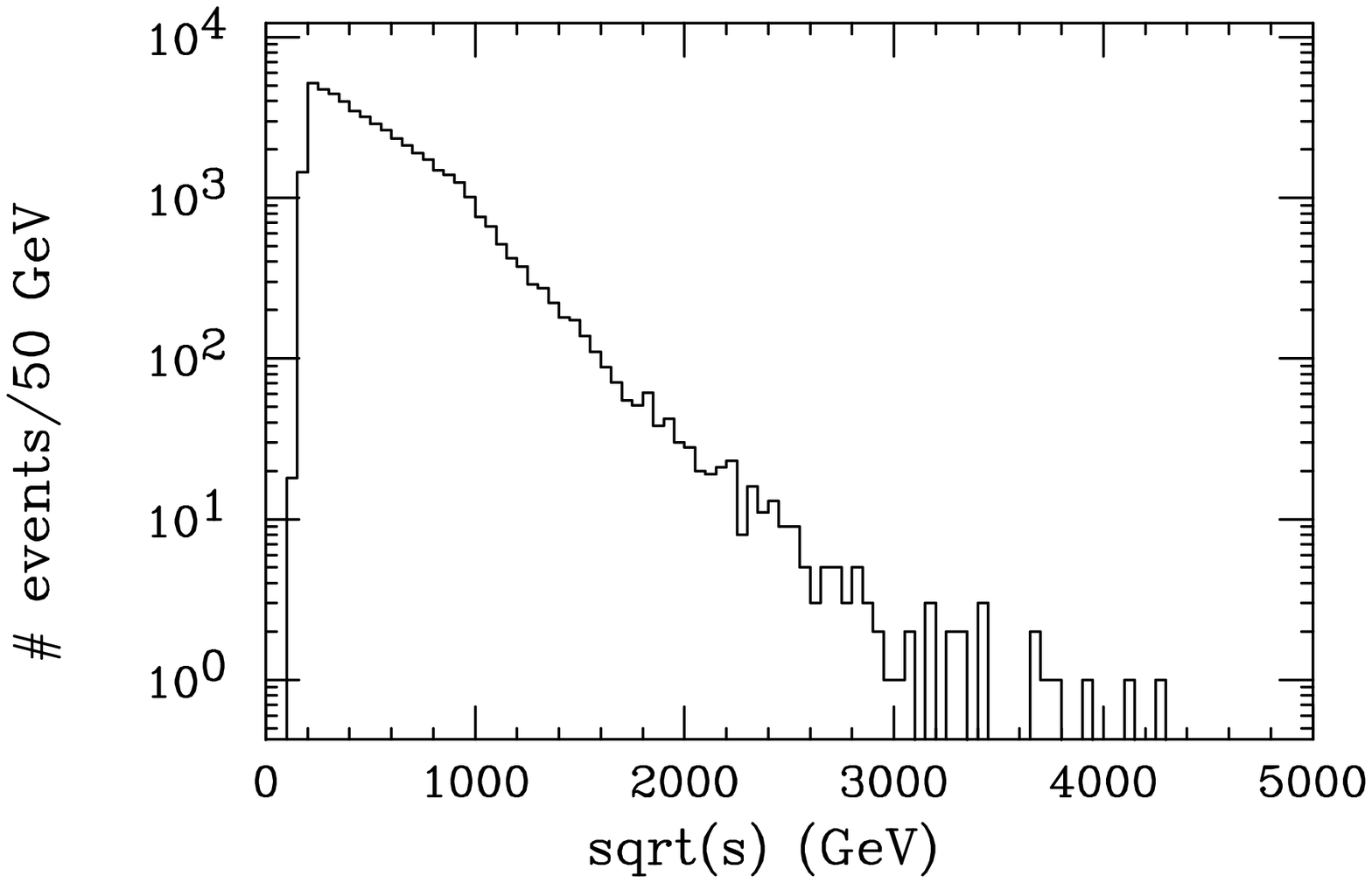}
\caption{Sample center of mass energy $\sqrt{s}$ distributions of $pp \to W^* \to WZ$ (left) and $pp \to h \to ZZ$ (right) in which collider sensitivities of TBVs are usually studied in previous literatures.}
\label{momdep2}
\end{figure}

In this paper, we use on-shell (constant) couplings rather than considering full momentum dependence. This momentum dependence comes from integrating out dynamical degrees of freedom, and are shown in \Fig{momdep}. Couplings around the threshold region are different from on-shell couplings. However, the typical energy scales of LHC processes that care applicability in the measurement of TBVs are only about $200 \GeV$ as shown in \Fig{momdep2}. The on-shell coupling thus may contribute more to the cross-section support than the threshold behavior. \Fig{momdep} also shows that the maximum couplings in the threshold region are only ${\cal O}(1)$ factor larger than the on-shell couplings. Although the threshold behavior depends on input parameters, we checked that maximum couplings are larger than the on-shell couplings by at most ${\cal O}(10)$ factor which does not affect our conclusion. It is also convenient to use on-shell couplings since it facilitates the comparison of our result with previous collider studies of TBVs that usually assume constant couplings.

\section*{Appendix B - Electric Dipole Moments}
EDM is a parity and time-reversal violating electromagnetic property of a fermion at the fermion mass scale. In field theory language, EDM comes from the CP-odd low-energy effective operator $-i\frac{1}{2} \bar{f} \sigma_{\mu\nu} \gamma_5 f F^{\mu\nu}$ with on-shell fermion $f$ and a photon. Exact full two-loop calculations have been carried out in \cite{Giudice:2005rz,Chang:2005ac,Feng:2008nm} for split supersymmetry, and in \cite{Pilaftsis:2002fe,Li:2008kz,West:1994} for the MSSM with one-loop EDM suppressed. In this appendix, we rather compute eEDM in split supersymmetry by inserting effective CP-odd TBVs into relevant diagrams in Fig.~\ref{EDM:one-two}. We work in dimensional regularization and $\overline{MS}$-scheme. It is a good way to check the previously computed results. For more accurate numerical analysis, we use the full two-loop results.

For reference, we list the leading order EDM in split supersymmetry (in the limit $M_1,M_2,\mu \gg M_W, M_h$) calculated using effective couplings.
\bea
d_f^{WW} &=& -\frac{e \alpha^2 T_f}{8 \pi^2 s_W^4} \sum_{i,k} \frac{m_f m_i^+ m_k^0}{M_W^2} \Im(C^{*L}_{ki} C^R_{ki}) \nonumber\\ && \qquad \cdot \frac{1}{m_i^2-m_k^2} \left( \frac{m_k^2}{m_i^2-m_k^2} \ln \frac{m_k^2}{m_i^2} + 1 \right) \cdot \left( \log \frac{\mu^2}{M_W^2} + \frac{3}{2} \right) \\
d_f^{\gamma h} &=& \frac{e Q_f \alpha^2}{4 \sqrt{2} \pi^2 s_W^2} \sum_i \Im( D_{ii}^R) \frac{m_f }{M_W m_i^+} \left( \frac{1}{2} \log \frac{\mu^2}{M_h^2} + \frac{3}{4} \right) \label{rh}\\
d_f^{Z h} &=& -\frac{e \alpha^2 (T^3_f - 2Q_f s_W^2)}{8 \sqrt{2} c_W^2 \pi^2 s_W^4} \sum_{i,j} \frac{m_f m_i^+}{M_W} \Im( D_{ij}^R F_{ji}^R - D_{ij}^L F_{ji}^L ) \nonumber\\
&& \qquad \cdot \frac{1}{m_i^2-m_j^2} \left( 1- \frac{m_j^2}{m_i^2 - m_j^2} \log \frac{m_i^2}{m_j^2} \right) \cdot \frac{1}{2} \left(\log \frac{\mu^2}{M_h^2} + \frac{M_Z^2}{M_h^2-M_Z^2} \log \frac{M_Z^2}{M_h^2} \right)
\eea
where superscripts imply two particles that mediate CP-violation to SM fermions. When two inos running in the loop are (almost) degenerate, these formula simplify as following.
\bea
d_f^{WW} &=& -\frac{e \alpha^2 T_f}{8 \pi^2 s_W^4} \sum_{i,k} \frac{m_f m_k^0}{M_W^2 m_i^+} \Im(C^{*L}_{ki} C^R_{ki}) \cdot \left( \log \frac{\mu^2}{M_W^2} + \frac{3}{2} \right) \\
d_f^{Z h} &=& -\frac{e \alpha^2 (T^3_f - 2Q_f s_W^2)}{8 \sqrt{2} c_W^2 \pi^2 s_W^4} \sum_{i,j} \frac{ m_f}{M_W m_i^+} \Im( D_{ij}^R F_{ji}^R - D_{ij}^L F_{ji}^L ) \nonumber\\ && \qquad \qquad \qquad \qquad \qquad \qquad \qquad \qquad \cdot \frac{1}{2} \left(\log \frac{\mu^2}{M_h^2} + \frac{M_Z^2}{M_h^2-M_Z^2} \log \frac{M_Z^2}{M_h^2} \right)
\eea
We checked that our results agree with the most recent calculations of~\cite{Li:2008kz}.

Effective matching scale $\mu$ may be chosen to obtain the EDM numerically close to the full two-loop result~\cite{ArkaniHamed:2004yi,Giudice:2005rz}:
\beq
\mu^2 = m_{\chi_1^+} m_{\chi_2^+}, \quad m_{\chi_1^+} m_{\chi_2^+}, \quad m_{\chi_1^+} m_{\chi_4^0} \qquad \textrm{for $\gamma h, Z h, WW$ respectively}
\eeq
We used the following relations, which follow from unitarity and the definitions of mixing matrices, to reach the final form:
\bea
\Im(D_{ij}^R F_{ji}^R) m_i^+ &=& \Im(D_{ji}^R F_{ij}^R) m_j^+ \qquad \textrm{no sum} \nonumber\\
\Im(D_{ij}^R F_{ji}^L) m_j^+ &=& - \Im(D_{ij}^L F_{ji}^L) m_i^+ \qquad \textrm{and $R \leftrightarrow L$.}
\eea

\end{document}